\documentclass[epj]{svjour}

\usepackage{graphicx}
\usepackage{fancyhdr}
\usepackage{amssymb}

\def\makeheadbox{{
\hbox to0pt{\vbox{\baselineskip=10dd\hrule\hbox
to\hsize{\vrule\kern3pt\vbox{\kern3pt
\hbox{\bfseries\@journalname\ manuscript No.}
\hbox{(will be inserted by the editor)}
\kern3pt}\hfil\kern3pt\vrule}\hrule}%
\hss}
}}
\begin{document}
\title{Complementarity of the CERN Large Hadron Collider and the \boldmath{$e^+e^-$}
       International Linear Collider\thanks{Extended version
       of Ref.\,\cite{Choi:2007zg} to be published in ``Supersymmetry on
       the Eve of the LHC", a special volume of European Physical Journal
       C, Particles and Fields (EPJC) in memory of Julius Wess.}}
\author{S.~Y.~Choi\thanks{\emph{Email: sychoi@chonbuk.ac.kr}}}
\institute{Department of Physics and RIPC, Chonbuk National University,
           Jeonju 561-756, Korea}
\date{\today}

\abstract{The next-generation high-energy facilities, the CERN Large Hadron
          Collider (LHC) and the prospective $e^+e^-$ International Linear Collider (ILC),
          are expected to unravel new structures of matter and forces
          from the electroweak scale to the TeV scale. In this report we
          review the complementary role of LHC and ILC in drawing a comprehensive
          and high-precision picture of the mechanism breaking the electroweak
          symmetries and generating mass, and the unification of forces in the
          frame of supersymmetry.
\PACS{{12.60.-i} {Models beyond the standard model} --
      {12.60.Jv} {Supersymmetric models}}
}

\maketitle

\section{Introduction}
\label{intro}

Particle physics has been very successful in unraveling the basic laws of
nature at the smallest accessible length scale, and it has revealed
a consistent picture, the Standard Model (SM), adequately describing the
structure of matter and forces, although the elusive Higgs boson is yet to be
identified \cite{Amsler:2008zz}. However, many theoretical arguments and
experimental observations indicate that the model is incomplete and that
it should be embedded in a more fundamental theory, addressing a set of crucial
questions to be approached experimentally at the TeV scale (Terascale): the
mechanism of electroweak symmetry breaking (EWSB) and mass generation; the
unification of forces, including gravity finally; and the structure of spacetime
at short distances.
This set of questions in particle physics is intriguingly connected to cosmology
questions such as the nature of particles comprising cold dark matter (CDM) and
the origin of the baryon asymmetry in the universe.

The next generation of high-energy accelerators will get access to the
Terascale with high expectation of providing decisive answers to these
questions \cite{R0a,R0b}. LHC with a c.m. energy of 14 TeV \cite{R1b,R1c}
will put the first springboard in 2008 for breakthrough discoveries in the
EWSB sector and in physics beyond the SM (BSM). However,
the analysis of new physics processes at LHC is complicated. Therefore,
an $e^+e^-$ facility with clean environments (and, potentially, with various
options such as $\gamma e$ and $\gamma\gamma$ collision modes and the GigaZ
mode running at an energy on top of the $Z$-boson resonance)
is required to complement this hadron machine
in drawing a comprehensive and high--resolution picture of EWSB and of the BSM.
The ILC \cite{R2a,R2b,R2c,R2d,R2e,R2f},
which is now in the design phase, would be
an excellent counterpart to LHC. The ILC energy of 500 GeV in the first phase
and 1 TeV in the upgraded phase in the lepton sector is equivalent in many
aspects to the higher LHC energy of about 5 TeV in the quark sector.
Moreover, ILC covers one of the most crucial energy ranges including the
characteristic EWSB scale $v=246$ GeV. [If the BSM scale revealed
at LHC might be beyond the reach of ILC, it could be accessed later by
the 3-5 TeV Compact Linear Collider (CLIC) \cite{R3}.]

Several dedicated studies of the interplay between LHC and ILC
have been carried out \cite{R2e,R4a,R4b}
in the recent past. In particular, the LHC/ILC Study Group, formed as
a collaborative effort of the hadron and lepton collider experimental
communities and theorists, has completed a comprehensive working group
report with detailed studies of various conceivable BSM scenarios \cite{R4a}.
Our report will not cover all these topics but it should give
a concise review of the complementary role of LHC and ILC in
drawing a model-independent and high-resolution picture of the new Terascale
physics which may reveal the fundamental theory at scales close to the grand
unification (GUT) or Planck scale. Supersymmetry (SUSY) will exclusively be
considered as a BSM prototype concept in this description.

\section{The supersymmetry path}
\label{sec:2}

In supersymmetric theories a light Higgs boson is generated and the electroweak
(EW) scale is stabilized naturally in the GUT/Planck-scale background.
The presence of the supersymmetric particle spectrum is essential
for the unification of the three SM gauge couplings at high energies
\cite{R6a,R6b,R6c,R6d}. It offers a natural CDM candidate. Moreover, local
SUSY provides a rationale for gravity by demanding the existence of massless
spin-2 gravitons. In short, if realized in nature, SUSY will have an impact
across all microscopic and cosmological scales.

\begin{figure}
\centering
\includegraphics[width=0.40\textwidth,height=0.43\textwidth,angle=0]{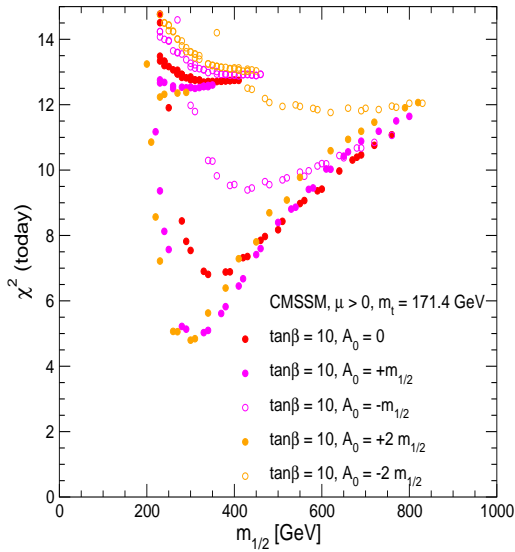}
\caption{The combined $\chi^2$ function for electroweak precision observables
         and $B$-physics observables as a function of the universal gaugino
         mass $m_{1/2}$; Ref.\,\cite{R9}.}
\label{fig1}
\end{figure}

\begin{figure}
\centering
\includegraphics[width=0.40\textwidth,height=0.39\textwidth,angle=0]{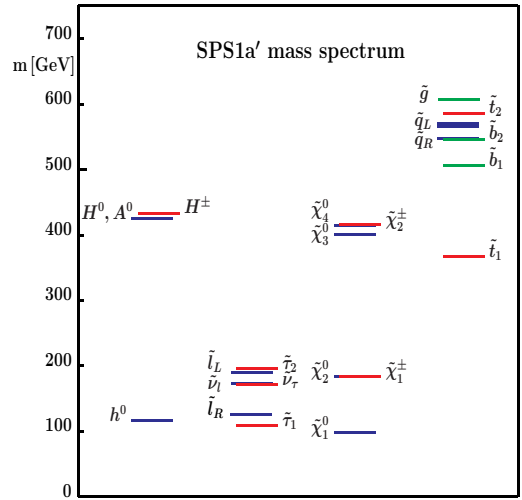}
\caption{Mass spectrum of supersymmetric particles and Higgs bosons in the
         reference point SPS1a$^\prime$; Ref.\,\cite{R10}.}
\label{fig2}
\end{figure}

There is no firm prediction for the SUSY mass scale. However,
direct bounds on the mass scale due to the absence of
sparticles at LEP and the Tevatron have been established, and important
indirect constraints
from the LEP lower limit of 114 GeV on the Higgs mass \cite{Higgs_bound},
the observation of photonic $b$-decays, $b\to s \gamma$ \cite{bsr1,bsr2},
the BNL measurement of the anomalous magnetic moment of the muon
$a_\mu$ \cite{R7b}, and also from the measurement of the CDM density
at WMAP \cite{R8} have been derived.
As shown in Fig.\,\ref{fig1}, a global fit to precision EW
and $B$-decay observables indicates a fairly low mass-spectrum for moderate
values of the Higgs mixing parameter $\tan\beta$ in minimal
supergavity (mSUGRA) \cite{R9}. In this favorable case several non-colored
supersymmetric particles such as lighter neutralinos and sleptons should be
observed at ILC in the first phase with 500 GeV c.m. energy and even the
heavier non-colored particles and the lighter top squark in the upgraded
phase with the c.m. energy of 1 TeV. The spectrum corresponding to a parameter
set with close to maximal probability is depicted in Fig.\,\ref{fig2}.
This spectrum had been chosen as a benchmark set for an mSUGRA scenario in
the SPS1a$^\prime$ project \cite{R10}.

LHC and ILC can provide us with a perfectly combined tool for exploring
SUSY \cite{R4a}. The heavy colored supersymmetric particles, squarks and
gluinos, can be discovered for masses up to 3 TeV with large rates at LHC.
The properties of the potentially lighter non-colored particles, charginos,
neutralinos, sleptons and Higgs bosons, can be studied very precisely at
ILC by exploiting, in particular, beam polarizations \cite{MoortgatPick:2005cw}.
Once the properties of the light particles are determined precisely at ILC,
the heavier particles produced at LHC can subsequently be studied in the cascade
decays with much greater precision. Based on the coherent LHC and ILC
analyses we can then take the supersymmetry path by
\begin{itemize}
\item[{$\bullet$}] measuring the masses and mixings of the newly produced
      particles, their decay widths and branching ratios, their production
      cross sections, etc;
\item[{$\bullet$}] verifying that there are indeed the superpartners of the SM
      particles by determining their spin and parity, gauge quantum numbers
      and their couplings;
\item[{$\bullet$}] reconstructing the low  energy Lagrangian with the smallest
      number of assumptions, i.e. as model independently as possible;
\item[{$\bullet$}] and unraveling the fundamental SUSY breaking mechanism and
      shedding light on the physics at the very high energy (GUT or Planck) scale,
\end{itemize}
from the EW scale to the GUT/Planck scale -- on one side, for the reconstruction
of the fundamental SUSY theory near the Planck scale and, on the other side, for
the connection of particle physics with cosmology.

\section{Higgs bosons}
\label{sec:3}

In SUSY theories the Higgs sector includes at least two iso-doublet
scalar fields so that at least five more physical particles are
predicted \cite{Higgs_LHC}. In the minimal supersymmetric SM (MSSM) the mass of
the lightest neutral scalar Higgs particle $h$ is bounded from above to about 140 GeV,
while the masses of the heavy neutral scalar and pseudoscalar Higgs bosons,
$H$ and $A$, and the charged Higgs bosons, $H^\pm$, may range from the EW
scale to the multi-TeV scale. The upper bound on the lightest Higgs mass is relaxed
only up to about 200 GeV in more general scenarios provided the fields remain
weakly interacting up to the Planck scale \cite{Espinosa:1998re}.

\subsection{The MSSM Higgs Bosons}
\label{sec:3-1}

\begin{figure}
\centering
\includegraphics[width=0.45\textwidth,height=0.45\textwidth,angle=0]{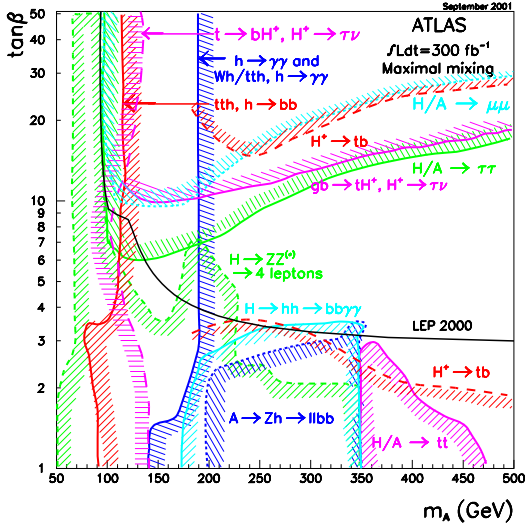}
\caption{The ATLAS sensitivity for the discovery of the MSSM Higgs bosons in the
         case of maximal mixing. The 5$\sigma$ discovery curves are shown in the
         $(\tan\beta,m_A)$ plane for the individual channels and for an integrated
         luminosity of 300 fb$^{-1}$. The corresponding LEP limit is also shown;
         Ref.\,\cite{richter}.}
\label{fig3}
\end{figure}

While the light Higgs boson $h$ can be detected at LHC in the full range of the
$M_A$ and $\tan\beta$ parameter space, none of the heavy Higgs bosons can be detected
in a blind wedge centered around the medium mixing angle $\tan\beta\sim 7$ and
opening from masses of about 200 GeV to higher values, cf.
Fig.\,\ref{fig3} \cite{richter,Heavy_Higgs}.
This region can be covered by ILC and CLIC up to the multi-TeV range.

\begin{figure}
\centering
\includegraphics[width=0.43\textwidth,height=0.46\textwidth,angle=0]{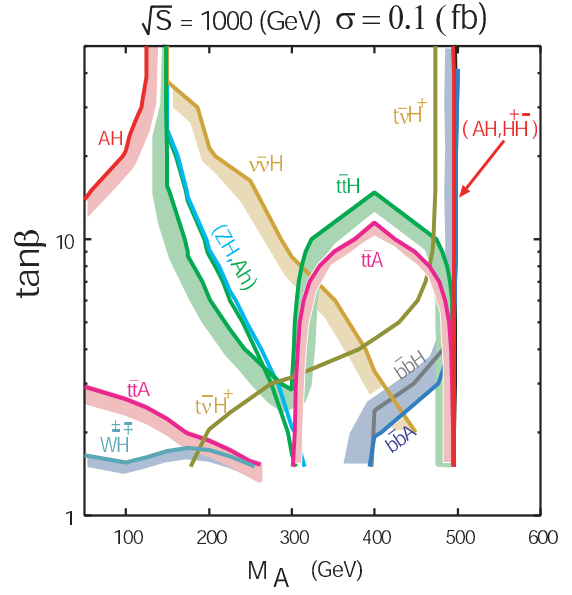}
\caption{Cross section contours of various heavy MSSM Higgs production
         processes in the $M_A/\tan\beta$ plane for $\sqrt{s}=1$ TeV;
         Ref.\,\cite{R11}.}
\label{fig4}
\end{figure}

At ILC the search and study of the light Higgs boson $h$ follows the pattern
very similar to the SM Higgs boson in most of the parameter space and the
heavy Higgs bosons are produced in mixed pairs at ILC:
$ e^+e^-\to HA\ \ \mbox{and}\ \ H^+H^-$.
Therefore, the wedge can be covered by pair production in $e^+e^-$ collisions
for masses $M_{H,A}\leq \sqrt{s}/2$, i.e., up to 500 GeV in the TeV phase of
the ILC machine, cf. Fig.\,\ref{fig4} \cite{R11} and, further, up to 2.5 TeV
at the 5 TeV CLIC \cite{R3}. Moreover, single production in photon-photon
collisions, $\gamma\gamma \to H$ and $A$, can cover the wedge up to Higgs
masses of 800 GeV if a fraction of 80\% of  the total energy of the
1 TeV ILC is transferred to the $\gamma\gamma$ system by
Compton back-scattering of laser light \cite{R12}. The mass reach for the heavy
Higgs bosons in $\gamma\gamma$ collisions can further be extended to 4 TeV at
the 5 TeV CLIC \cite{Krawczyk:2008cg}.

After the Higgs bosons are discovered, it must experimentally be established
that the Higgs mechanism is responsible indeed for breaking the EW symmetry
and for generating the masses of the fundamental SM particles. This
requires the precise profiling of the Higgs bosons. First model-independent analyses
of the properties can be performed at LHC by measuring the Higgs masses,
the Higgs spin(s) \cite{Choi:2002jk,Buszello:2002uu}, the ratios of some Higgs
couplings and the bounds on couplings \cite{R13}.

\begin{figure}[h]
\centering
\includegraphics[width=0.43\textwidth,height=0.50
                 \textwidth,angle=0]{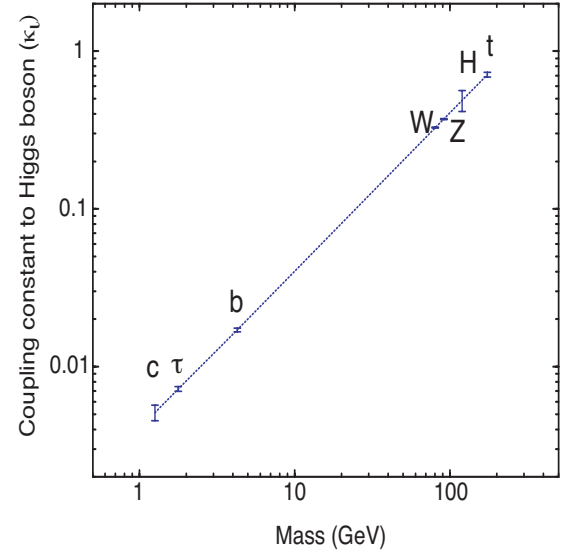}
\caption{The relation between the Higgs coupling of a particle and its mass in
         the SM. The error bars correspond to the accuracy expected from ILC
         data; Ref.\,\cite{GLC_Higgs_coupling_mass_relation}.}
\label{fig5}
\end{figure}

However, the truly model-independent and  high-resolution determination of the
profile of the light Higgs boson $h$ -- the mass, the spin of the
particle, the absolute values of the Higgs couplings to the SM particles and the
trilinear Higgs self couplings --  can be carried out at ILC, with clear
signals of Higgs events above small backgrounds in the processes such as
Higgs--strahlung, $e^+e^-\to Zh$, and $WW$ fusion, $e^+e^-\to \bar{\nu}\nu h$,
and in the process of double Higgs production, $e^+e^-\to Zhh$ and
$\bar{\nu}\nu hh$ \cite{Djouadi:1996ah}. As shown in Fig.\,\ref{fig5}
for typical SM particle species,
the linear relation between the Higgs couplings and the masses for typical
SM particle species can be tested
with great precision at ILC \cite{GLC_Higgs_coupling_mass_relation}.

\begin{figure}[h]
\centering
\includegraphics[width=0.40\textwidth,height=0.40\textwidth,angle=0]{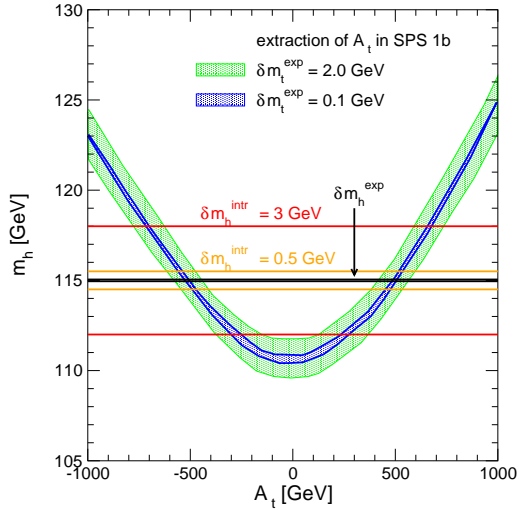}
\caption{Extracting the trilinear coupling $A_t$ from radiative corrections to
         the light MSSM Higgs mass; Ref.\,\cite{R14}.}
\label{fig6}
\end{figure}

High-precision measurements of the lightest Higgs mass at ILC can be exploited to
determine parameters in the SUSY theory which are difficult to
measure directly. For instance, by evaluating quantum
corrections, the top quark trilinear coupling $A_t$ can be calculated from the
Higgs mass, Fig.\,\ref{fig6}. For an error on the top quark mass of
$\delta m_t=100$ MeV and an error on the Higgs mass of $\delta m_h=50$ MeV,
$A_t$ can be determined at an accuracy of about 10\% \cite{R14}.

\begin{figure}[h]
\centering
\includegraphics[width=0.40\textwidth,height=0.40\textwidth,angle=0]{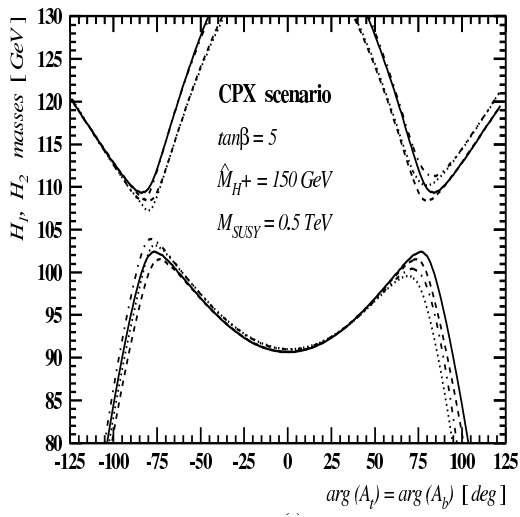}
\caption{Numerical estimates of the $H_{1,2}$ pole masses as a function of
        the CP-violating phase of the stop/sbottom trilinear parameter $A_{t,b}$;
        Ref.\,\cite{R18}.}
\label{fig7}
\end{figure}

\subsection{CP violation in the MSSM Higgs sector}
\label{sec:3-2}

In the general MSSM \cite{general_MSSM}, the gaugino mass parameters
$M_i$ ($i=1,2,3$), the higgsino mass parameter $\mu$, and the trilinear
couplings $A_f$ can be complex so that they can induce explicit CP violation
in various ways in the model. Their physical combinations
affect sparticle
masses and couplings through mixing, induce CP-violating mixing in the
Higgs sector through radiative corrections, influence CP-even observables such
as cross sections and also lead to interesting CP-odd asymmetries at colliders.
As a result, although stringently constrained by low energy observables like
electric dipole moments (EDMs), the nontrivial CP phases can significantly
influence the collider phenomenology of Higgs and SUSY particles and also the
properties of neutralino CDM \cite{R15a,R15b,R15c}.

\begin{figure}[h]
\centering
\includegraphics[width=0.45\textwidth,height=0.47\textwidth,angle=0]{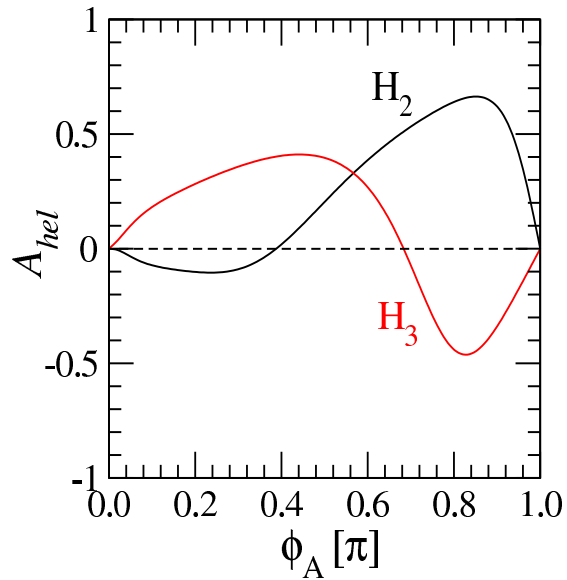}
\caption{The CP-odd asymmetry $A_{hel}$ at the pole of the heavy Higgs bosons
         $H_2$ and $H_3$ as a
         function of the phase $\phi_A$ of the stop trilinear
         parameter $A_t$; Ref.\,\cite{R17}.}
\label{fig8}
\end{figure}

Referring to the CPNSH report \cite{R15a} for an extensive discussion of CP
violation in supersymmetric theories, we mention in this report just two examples
of CP-violation in the Higgs-sector. The lightest Higgs boson $H_1$ without definite
CP-parity may couple very weakly to the gauge bosons so that the state could have
escaped detection at LEP2 \cite{R16}, and the heavy Higgs states $H$ and $A$ can
exhibit CP-violating resonant mixing phenomena when two states are degenerate
in mass in the decoupling regime \cite{R17}. One example of the impact of
the CP-violating Higgs mixing on the Higgs mass spectrum is shown in
Fig.\,\ref{fig7} as a function of the phase of the stop and sbottom
trilinear coupling $A_{t,b}$ \cite{R18}.
The other example for studying the CP-violating resonant mixing of
two heavy neutral Higgs bosons is provided by $\gamma\gamma$-Higgs formation
in polarized beams. As shown in Fig.\,\ref{fig8}, the CP violation due to
resonant $H/A$ mixing can directly be probed via the CP-odd asymmetry
$A_{hel}=(\sigma_{++}-\sigma_{--})/(\sigma_{++}+\sigma_{--})$ constructed with
circular photon polarization \cite{R17}.

\subsection{Extended Higgs sector}
\label{sec:3-3}

A large variety of BSM theories such as GUT theories and string theories suggest
extended gauge and Higgs sectors with additional gauge bosons and Higgs bosons
beyond the minimal set of the MSSM \cite{R19a,R19b,R19c,R19d,R37,R20}.

\begin{figure}
\centering
\includegraphics[width=0.50\textwidth,height=0.50
                 \textwidth,angle=0]{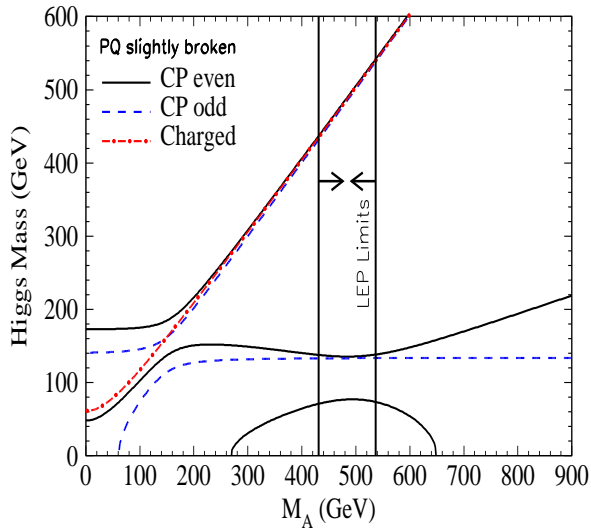}
\caption{Typical Higgs mass spectrum in the NMSSM; Ref\,\cite{R19a}.}
\label{fig9}
\end{figure}

The next-to-MSSM (NMSSM), the simplest extension of the MSSM,
introduces a complex iso-scalar field, generating a weak scale higgsino mass
parameter $\mu$ by spontaneous symmetry breaking in the Higgs sector.
The NMSSM Higgs sector is thus extended to include an additional
scalar and a pseudoscalar. The axion-type character of the pseudoscalar
boson renders this particle preferentially light. An example for the mass
spectrum \cite{R19a} is shown in Fig.\,\ref{fig9}. Since the trilinear couplings
increase with energy, upper bounds on the mass of the lightest neutral scalar
Higgs boson can be derived from the assumption that the theory be valid
up to the GUT scale: $m(H_1)\lesssim 140$ GeV. Thus, in spite of the additional
interactions, the distinct pattern of the minimal extension remains valid
also in more complex supersymmetric scenarios. If $H_1$ is (nearly) pure
isosinglet, the coupling $ZZH_1$ is small and the particle cannot be
produced by Higgs-strahlung. However, in this case $H_2$ is generally light
and couples with sufficient strength to the $Z$ boson; if not, $H_3$
plays this role, so that one Higgs boson can be discovered in any case.

A large variety of other extensions beyond the SM has been analyzed
theoretically. For instance, if the gauge boson sector is expanded by an
additional U(1)$'$ Abelian symmetry at high energies \cite{R37}, the
additional pseudoscalar Higgs field is absorbed to generate the mass of the
new $Z'$ boson while the scalar part of the Higgs field can be observed
as a new Higgs boson beyond the MSSM set. If generated by an extended
symmetry like $E_6$, the Higgs sector is expanded by an ensemble
of new states \cite{R20} with quite unconventional properties.

Quite generally, as long as the fields in supersymmetric theories remain
weakly interacting up to the canonical Planck scale, the mass of the lightest
Higgs bosons is bounded to about 200 GeV as the Yukawa couplings are
restricted to be small in the same way as the quartic coupling in the standard
Higgs potential. Moreover, the mass bound of 140 GeV for the lightest Higgs
particle is realized in almost all supersymmetric theories \cite{Espinosa:1992hp}.
Consequently, experiments at ILC with 500 GeV c.m. energy are in a
{\it no-lose} situation \cite{Kamoshita:1994iv} for detecting the Higgs particles
even in general supersymmetric theories.

\section{Supersymmetric particles}
\label{sec:4}

For an explicit numerical illustration we adopt the parameters of the minimal
supergravity reference point SPS1a$^\prime$ \cite{R10}. It is characterized by the
following values of the soft parameters at the GUT scale: $M_{1/2}=250$ GeV,
$M_0=70$ GeV, $A_0=-300$ GeV, ${\rm sign}(\mu)=+ $ and $\tan\beta=10$ where
$M_{1/2}$, $M_0$, $A_0$ and $\mu$ denote the universal gaugino mass, the universal
scalar mass, the universal trilinear coupling and the higgsino mass parameter.
The modulus of the higgsino mass parameter is fixed by requiring radiative electroweak
symmetry breaking \cite{EWSB1,EWSB2,EWSB3,EWSB4,EWSB5} so that $\mu=+396$ GeV.
As shown by the sparticle and Higgs spectrum in Fig.\,\ref{fig2}, the squarks and
gluinos can be studied very well at LHC and the non-colored charginos and
neutralinos, sleptons and Higgs bosons can be analyzed partly at LHC and studied precisely
at ILC operating at a c.m. energy up to 1 TeV.

\subsection{Masses of supersymmetric particles}
\label{sec:4-1}

At LHC, the masses can be obtained by analyzing edge effects in the cascade
decay spectra, cf.\,Ref.\,\cite{R22}. An ideal chain is a long sequence of
two-body decays: $\tilde{q}_L\to\tilde{\chi}^0_2 q \to \tilde{\ell}_R \ell q \to
\tilde{\chi}^0_1\ell\ell q$. The kinematic edges and thresholds predicted
in the invariant mass distributions of the two leptons and the jet determine
the masses in a model-independent way \cite{R22}. The four particle masses
measured by this method
are used subsequently as input for other decay chains like $\tilde{g}\to
\tilde{b}_1b\to\tilde{\chi}^0_2 bb$ and the shorter chains
$\tilde{q}_R\to q\tilde{\chi}^0_1$ and $\tilde{\chi}^0_4\to\tilde{\ell}\ell$.
However, there are residual ambiguities and the strong correlations between
the heavier masses and the lightest supersymmetric particle (LSP).

\begin{figure}
\centering
\includegraphics[width=0.44\textwidth,height=0.43\textwidth,angle=0]{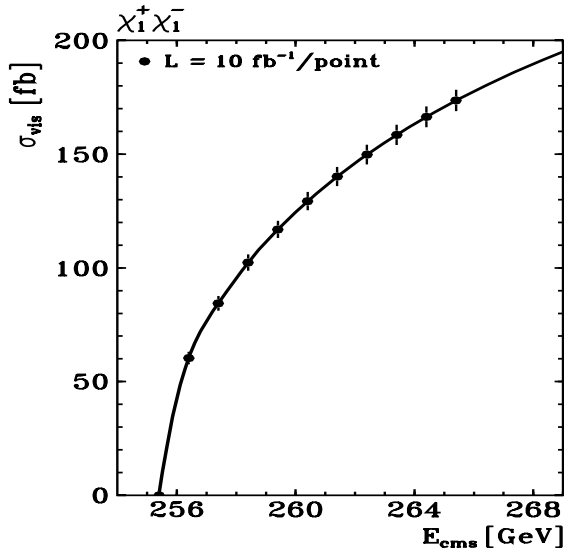}
\caption{Mass measurement at the threshold of chargino $\tilde{\chi}^+_1\tilde{\chi}^-_1$
         pair production; Ref\,\cite{R23}.}
\label{fig10}
\end{figure}
\begin{figure}
\centering
\includegraphics[width=0.43\textwidth,height=0.43\textwidth,angle=0]{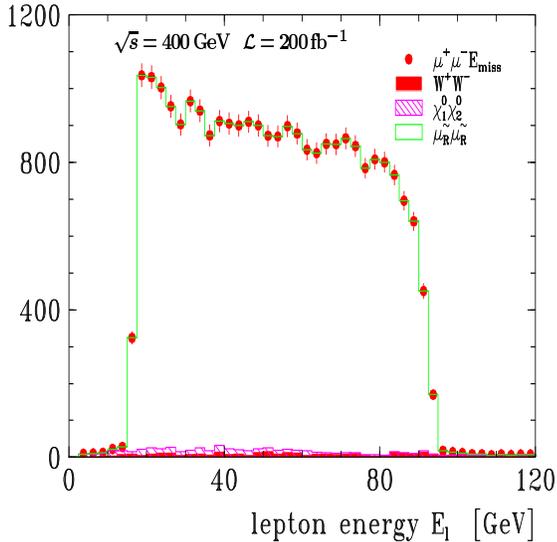}
\caption{Smuon and neutralino edges in two-body smuon decays,
         $\tilde{\mu}^\pm_R\to\mu^\pm\tilde{\chi}^0_1$;
         Ref\,\cite{R23}.}
\label{fig11}
\end{figure}
\begin{table}
\caption{Accuracies of representative mass measurements of SUSY particles in
         individual LHC, ILC and coherent LHC/ILC analyses in the point
         SPS1a$'$ [mass units in GeV]; Ref.\,\cite{R10}.}
\label{tab:1}
\vskip 0.3cm
\begin{center}
\begin{tabular}{|c|c||c|c||c|}
\hline
Particles & Mass &  ``LHC"  & ``ILC"  & ``LHC+ILC" \\
\hline \hline
$h^0$                 & 116.0 & 0.25 & 0.05 & 0.05 \\
$H^0$                 & 425.0 &      & 1.5  & 1.5  \\
    \hline
$\tilde{\chi}^0_1$    &  97.7 & 4.8  & 0.05 & 0.05 \\
$\tilde{\chi}^0_2$    & 183.9 & 4.7  & 1.2  & 0.08 \\
$\tilde{\chi}^0_4$    & 413.9 & 5.1  & 3-5  & 2.5  \\
$\tilde{\chi}^\pm_1$  & 183.7 &      & 0.55 & 0.55 \\ \hline
$\tilde{e}_R$         & 125.3 & 4.8  & 0.05 & 0.05 \\
$\tilde{e}_L$         & 189.9 & 5.0  & 0.18 & 0.18 \\
$\tilde{\tau}_1$      & 107.9 & 5-8  & 0.24 & 0.24 \\ \hline
$\tilde{q}_R$         & 547.2 & 7-12 & -    & 5-11 \\
$\tilde{q}_L$         & 564.7 & 8.7  & -    & 4.9  \\
$\tilde{t}_1$         & 366.5 &      & 1.9  & 1.9  \\
$\tilde{b}_1$         & 506.3 & 7.5  & -    & 5.7  \\ \hline
$\tilde{g}$           & 607.1 & 8.0  & -    & 6.5  \\ \hline
\end{tabular}
\end{center}
\end{table}

At ILC very precise mass values can be extracted from threshold scans and decay
spectra \cite{R23}. The excitation curves for chargino $\tilde{\chi}^\pm_{1,2}$
production in S-waves rise steeply with the velocity of the particles near
threshold and they are thus very sensitive to the mass values, cf. Fig.\,\ref{fig10}.
The same holds true for mixed chiral selectron pairs in
$e^+e^-\to\tilde{e}^+_R\tilde{e}^-_L$ and for diagonal pairs in
$e^-e^-\to \tilde{e}^-_R\tilde{e}^-_R, \tilde{e}^-_L\tilde{e}^-_L$ \cite{R24a,R24b}.
Other scalar fermions as well as neutralinos are produced in P-waves
with a less steep threshold behavior proportional to the third
power of the velocity. Important information on the mass of the LSP
such as the lightest neutralino $\tilde{\chi}^0_1$ can be obtained
from the sharp edges of two-body decay spectra as $\tilde{\ell}_R
\to \ell \tilde{\chi}^0_1$, cf. Fig.\,\ref{fig11} \cite{R23}.
The accuracy in the measurement of the LSP mass can be improved at ILC by
two orders of magnitude compared with LHC; Tab.\,\ref{tab:1}.

The values of typical mass parameters and their related measurement errors are
presented in Tab.\,\ref{tab:1}: ``LHC" from LHC analyses and ``ILC" from
ILC analyses. The fourth column ``LHC+ILC" represents the corresponding
errors if the experimental analyses are performed coherently \cite{R10}.

\subsection{Spins of supersymmetric particles}
\label{sec:4-2}

Determining the spin of new particles is an important method to clarify the
nature of the particles and the underlying theory. This determination is
crucial to distinguish the supersymmetric interpretation of new particles
from other models.

The measurement of the spins in particle cascades at LHC
is quite involved \cite{R25a,R25b,R25c}. While the invariant mass distributions
of the particles in decay cascades are characteristic for the spins of the
intermediate particles involved, detector effects strongly reduce the
signal in practice.

\begin{figure*}
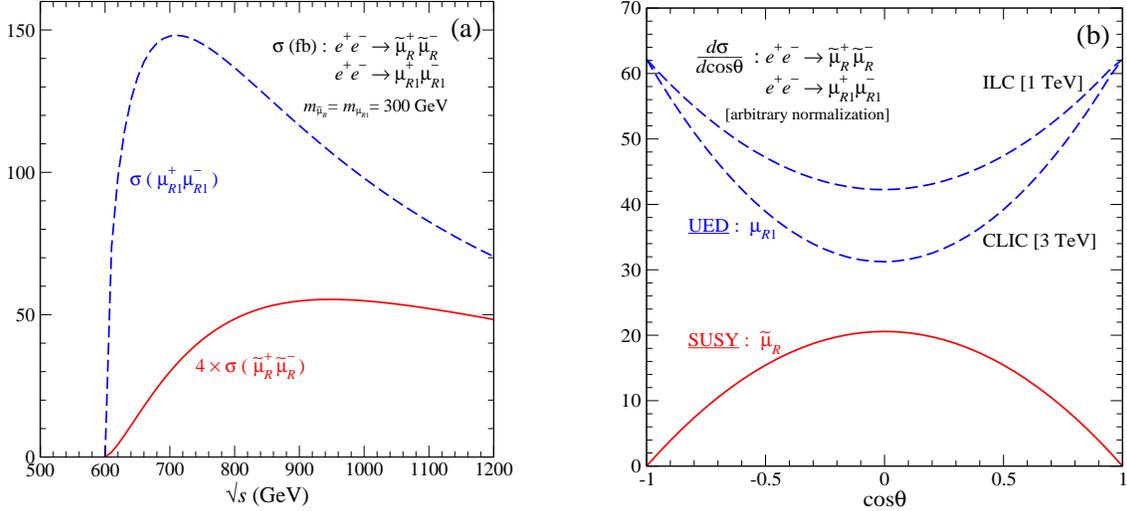

\centering
\includegraphics[width=0.4\textwidth,height=0.4
                 \textwidth,angle=0]{smuon_threshold.eps}
\hskip 1.3cm
\includegraphics[width=0.4\textwidth,height=0.4
                 \textwidth,angle=0]{smuon_angle.eps}
\caption{The threshold excitation (a) and the angular distribution (b)
         in the case of smuons in the MSSM, compared with the first
         spin-1/2 Kaluza-Klein muons in a model of universal extra
         dimensions, in pair production at ILC; for details,
         see Ref.\,\cite{R26b}.}
\label{fig12}
\end{figure*}

In contrast, the spin measurement at ILC is straightforward \cite{R26a,R26b}.
Even though the P-wave onset of the excitation curve is a necessary but not
sufficient condition, the $\sin^2\theta$ law for the angular distribution in the
production of sleptons (for selectrons close to threshold)
is a unique signature of the fundamental spin-zero character;
Fig.\,\ref{fig12}. On the contrary, neither the onset of the excitation curves near
threshold nor the angular distribution in the production processes provide unique
signals of the spin of charginos and neutralinos. However, decay angular
distributions of polarized charginos/neutralinos,
as generated naturally in $e^+e^-$ collisions, can provide an
unambiguous determination of the spin-1/2 character of the particles
albeit at the expense of more involved experimental analyses \cite{R26b}.
[Quantum interference among helicity amplitudes, reflected in azimuthal angle
distributions, may provide another method for determining
spins \cite{azimuthal}. However, this method depends strongly on the masses of
the decay products and the c.m. energy, as the quantum interference disappears
with increasing energy.]

\subsection{Mixings of supersymmetric particle states}
\label{sec:4-3}

\begin{figure}[h]
\centering
\includegraphics[width=0.43\textwidth,height=0.43\textwidth,angle=0]{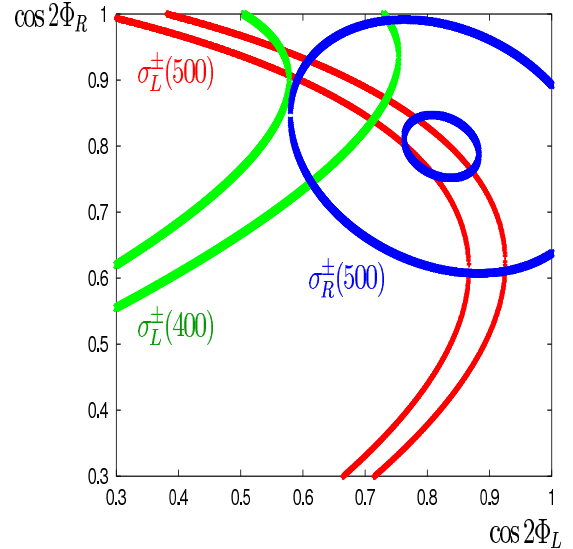}
\caption{Contours of mixing angles for the $e^+e^-\to\tilde{\chi}^+_1\tilde{\chi}^-_1$ production
         cross section for polarized $e^\pm$ beams at $\sqrt{s}=400$ and $500$
         GeV; Ref.\,\cite{R27c}.}
\label{fig13}
\end{figure}

Mixing parameters must be extracted from measurements of cross sections and
polarization asymmetries. [The determinations of mixing parameters are
difficult at LHC since several production and decay processes are simultaneously
involved and only the products of the production cross sections and the decay
branching fractions are measured experimentally.]
In the production of charginos and neutralinos,
both diagonal and non-diagonal pairs can be exploited: $e^+e^-$ $\to$ $\tilde{\chi}^+_i
\tilde{\chi}^-_j$ [$i,j=1,2$] \cite{R27a,R27b,R27c} and
$\tilde{\chi}^0_i\tilde{\chi}^0_j$ [$i,j=1,..,4$] \cite{R28}. The production
cross sections for charginos are binomials in $\cos 2\phi_{L,R}$ where
$\phi_{L,R}$ are the mixing angles rotating current to mass eigenstates. Using
polarized electron and positron beams, the mixings can be determined in a
model-independent way, Fig.\,\ref{fig13}. The same procedures can be applied to
determine the mixings in the sfermion sector \cite{R29a,R29b,Bartl:1997yi,R29c}.
The production cross sections for stop particle pairs,
$e^+e^-\to\tilde{t}_i \tilde{t}^*_j$ [$i,j=1,2$], depend on the
stop mixing angle $\theta_{\tilde{t}}$ which can be determined
with high accuracy by use of polarized electron beams \cite{Bartl:1997yi,R29c}.

\subsection{Supersymmetric Yukawa couplings}
\label{sec:4-4}

SUSY predicts the identity of Yukawa and gauge couplings among supersymmetric
partners for gauge bosons and gauginos, and for fermions and their scalar partners.

The identity of the SU(3) QCD Yukawa and gauge couplings can be studied
experimentally at LHC through pair production of squarks partly mediated by
gluino $t$-channel exchanges \cite{R30a,R30b}. A potential
complement \cite{Brandenburg:2008gd} to this method is gluino emission in
association with quark-squark final states in $e^+e^-$ collisions,
$e^+e^-$ $\to q\tilde{q}\tilde{g}$, which might be kinematically accessible at the
second phase of ILC and/or at CLIC.  While the $q\tilde{q}
\tilde{g}$ channel measures the $q\tilde{q}\tilde{g}$ Yukawa coupling, the
radiation processes $\tilde{q}\tilde{q} g$ and $qqg$ determine the
QCD gauge coupling in the squark sector and the standard quark sector for
comparison.

On the other hand the SU(2) weak  and U(1) hypercharge relations can be
confirmed experimentally at ILC through pair production of charginos and neutralinos
which is partly mediated by the exchange of sneutrinos and selectrons in the
$t$-channel \cite{R28}, as well as selectron and sneutrino production which is
partly mediated by neutralino and chargino exchanges \cite{R24b}.
The separation of the electroweak SU(2) and U(1) couplings is also possible
if polarized electron beams are available. Of course the analysis for confirming
the identity of Yukawa and gauge couplings should be
performed by taking into account the prior measurements of the masses and/or
mixing parameters of the particles exchanged in the $t$-channel.
Taking into account uncertainties from the selectron and the neutralino parameters,
the SU(2) and U(1) Yukawa couplings can be extracted with a precision of
0.7\% and 0.2\%, respectively, at ILC with a 500 GeV energy and 500 fb$^{-1}$
integrated luminosity in the SPS1a$'$ scenario.

\subsection{Majorana versus Dirac fermions}
\label{sec:4-5}

The parallelism between self-conjugate neutral vector gauge bosons and their
fermionic supersymmetric partners induces the Majorana nature of these
particles in the minimal formulation of the theory.  Nevertheless,
experimental tests of the Majorana character of gluinos and neutralinos
would provide non-trivial insight into the realization of SUSY in
nature, since extended supersymmetric models can include Dirac gauginos.
$N=2$ SUSY provides a solid theoretical basis for formulating such a testing
ground \cite{Benakli}. Since the fermionic degrees of freedom are doubled in
the gauge sector, the ensuing two Majorana fields can be joined to a single
Dirac field if the masses are chosen identical.

There are several methods to investigate the Majorana nature of gluinos
at LHC. In the original form, decays to heavy stop/top quarks are exploited
\cite{gluino_Majorana} to test whether the final state in the fermion decay $
\tilde{g} \to \tilde{t} \bar{t} + {\tilde{t}}^\ast t$ is self-conjugate.
The standard production processes for investigating the Majorana nature of
gluinos \cite{Choi:2008pi} are the production of a pair of equal-chirality
squarks, $q_Lq_L\to \tilde{q}_L\tilde{q}_L$ and $q_Rq_R\to\tilde{q}_R\tilde{q}_R$.
While the cross section for the scattering processes with
equal-chirality quarks is non-zero in the Majorana theory, it vanishes in the Dirac
theory. Owing to the dominance of $u$-quarks over $d$-quarks in the proton,
the Majorana theory predicts large rates of like-sign dilepton final states from
squark pair production with an excess of positively charged leptons while they
are absent, apart from a small number of remnant channels, in the Dirac theory.
[In a realistic analysis one has to include gluino production processes which can
also feed the like-sign dilepton signal but can be discriminated by extra jet
emission from the gluino decays.]

On the other hand, the Majorana/Dirac nature of neutralinos can be studied
through very clean reactions, $e^- e^- \to {\tilde{e}}^- {\tilde{e}}^-$, in the
$e^-e^-$ collision mode of ILC with left/right-handedly polarized
beams \cite{Choi:2008pi,Keung:1983nq,AguilarSaavedra:2003hw}.

\subsection{Split supersymmetry}
\label{sec:4-6}

For the unification of forces at the GUT scale the sfermion mass scale
$M_0$ is irrelevant, since each generation of sfermions furnishes a complete
SU(5) [or SO(10), if right-handed sneutrinos are included]. Likewise, the
dark-matter prediction of the MSSM and its extensions does not rely on the
value of $M_0$, but rather on the existence of a conserved discrete quantum
number, R-parity. These quantitative considerations led to the speculation
that the sfermion mass scale may actually be much higher than the gaugino
mass scale, effectively removing all scalar partners of the matter fields
and the extra heavy Higgs states of the MSSM from the low-energy
spectrum \cite{r37.1}.

With such a high sfermion mass scale, e.g., $M_0\sim 10^9$ GeV, the gluino
acquires a macroscopic lifetime and, for the purpose of collider experiments,
it behaves like a massive, stable color-octet parton. This leads to
characteristic signatures at LHC. However, due to the absence of cascade
decays, the production of the non-colored gauginos and higgsinos at LHC proceeds
only via EW annihilation processes, and the production rates are thus
considerably suppressed compared to conventional MSSM scenarios.

In this situation, the analysis of chargino and neutralino pair-production
at ILC provides the information necessary to deduce the supersymmetric nature
of the model \cite{Kilian:2004uj}. Extracting the values of chargino and
neutralino Yukawa couplings,
responsible for the mixing of gaugino and higgsino states, reveals the
anomalous effects due to the splitting of gaugino and sfermion mass scales.

\section{The fundamental theory}
\label{sec:5}

Combining the information from LHC on the generally heavy colored particles
with the information from ILC on the generally lighter non-colored particle
sector (and later from CLIC on heavier states) will generate a model-independent
and high-precision picture of SUSY at the Terascale. The picture may subsequently
serve as a solid platform for the reconstruction of the fundamental SUSY theory
at a high scale, potentially close to the Planck scale, and for the analysis
of the microscopic mechanism of SUSY breaking \cite{R31,Raby}. The experimental
accuracies expected at the per-cent down to the per-mil level must be matched on
the theoretical side. This demands a well-defined framework for the calculational
schemes in perturbation theory as well as for the input parameters like
a recently proposed scheme called Supersymmetry Parameter Analysis (SPA) \cite{R10}.

\subsection{Linking Terascale SUSY to the GUT/Planck scale}
\label{sec:5-1}

If SPS1a$^\prime$ or a similar SUSY parameter set is realized in nature, various
channels can be exploited to extract the basic Terascale SUSY parameters at LHC
and ILC. The data analysis performed coherently for LHC and ILC
gives rise to a very precise picture of the supersymmetric particle spectrum.
Running global analysis programs for the whole set of data enables us to
extract the Lagrangian parameters in the optimal way after including
radiative corrections \cite{R32a,R32b,R32c,R32d,R32e}.
The present quality of such an analysis can be judged from the results shown
in Tab.\,\ref{tab:2}.

\begin{table}
\caption{Excerpt of extracted SUSY Lagrangian mass and Higgs parameters at the
         Terascale in the reference point SPS1a$'$ [mass units in
         GeV]; Ref.\,\cite{R10}.}
\label{tab:2}
\vskip 0.3cm
\begin{center}
\begin{tabular}{|c|c|c|}
\hline
Parameter & SPS1a$'$ value &  Fit error [exp] \\
\hline \hline
$M_1$             & 103.3 & 0.1  \\
$M_2$             & 193.2 & 0.1  \\
$M_3$             & 571.7 & 7.8  \\
$\mu$             & 396.0 & 1.1  \\
    \hline
$M_{L_1}$    & 181.0 &  0.2 \\
$M_{E_1}$    & 115.7 &  0.4 \\
$M_{L_3}$    & 179.3 &  1.2 \\
   \hline
$M_{Q_1}$    & 525.8 &  5.2 \\
$M_{D_1}$    & 505.0 & 17.3 \\
$M_{Q_3}$    & 471.4 &  4.9 \\
   \hline
$m_A$          & 372.0  & 0.8 \\
$A_t$          & -565.1 & 24.6 \\
$\tan\beta$    & 10.0 & 0.2\\ \hline
\end{tabular}
\end{center}
\end{table}
\begin{figure*}[t]
\centering
\includegraphics[width=0.95\textwidth,height=0.40\textwidth,angle=0]{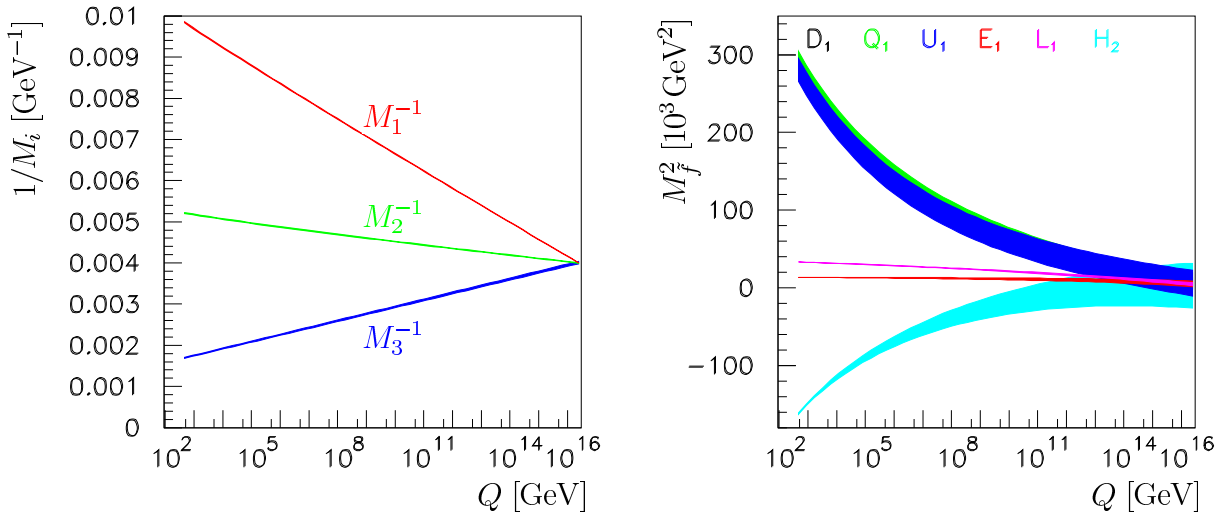}
\caption{The evolution of the gaugino and scalar mass parameters with the
         scale $Q$ in SPS1a$^\prime$. Only experimental errors are taken into
         account; theoretical errors are assumed to be reduced to the same size
         in the future; Ref.\,\cite{R31}.}
\label{fig14}
\end{figure*}

Based on the parameters extracted at the Terascale we can reconstruct the
fundamental SUSY theory and the related microscopic picture of the SUSY
breaking mechanism \cite{R31}. The experimental information is exploited
to the maximum extent possible in the bottom-up approach in which the
extrapolation from the Terascale to the GUT/Planck scale is performed
by the renormalization group (RGE) evolution of all parameters, with the
GUT scale defined by the unification point of the gauge couplings.

Typical examples for the evolution of the gaugino and scalar mass
parameters are presented in Fig.\,\ref{fig14}. While the determination
of the high-scale parameters in the gaugino and higgsino sector, as well
as in the non-colored slepton sector, is very precise, the picture of
the colored scalar and Higgs sectors is still coarse so that
efforts should be made to refine it considerably.
If the structure of the theory at the GUT scale were known a priori and merely
the experimental determination of the high scale parameters were lacking, then
the top down-approach would lead to a very precise parametric picture at the
Terascale.

So far, we have only considered the MSSM, in particular the parameter set
SPS1a$'$, as a benchmark scenario for judging the coherent capabilities of LHC
and ILC experiments for the analysis of future SUSY data. However,
neither this specific point nor the MSSM itself may be the correct model for
low-scale SUSY. Various extended models beyond the MSSM have therefore to be
investigated. The ILC experiments are crucial in discriminating between such
various theories beyond the SM as they enable us to measure low-scale
parameters directly with great precision and to estimate high-scale
parameters reasonably well.

\subsection{Left-right supersymmetric extension}
\label{sec:5-2}

A well-motivated example of model parameterizations at the very high scale,
different from the mSUGRA scenario, is provided by models
incorporating the right-handed neutrino sector to
accommodate the complex structure observed in the neutrino
sector. This requires the extension of the MSSM by a superfield including
the right-handed neutrino field together with its scalar partner. If the small neutrino
masses are generated by the seesaw mechanism, a similar type of spectrum
is induced in the scalar sneutrino sector, splitting into light TeV scale
and very heavy masses. The intermediate seesaw scales will affect the
evolution of the SUSY breaking soft mass terms at the high (GUT) scale,
especially in the third generation with large Yukawa couplings. This
provides us with the opportunity to measure, indirectly, the intermediate
seesaw scale of the third generation \cite{R33,Deppisch:2007xu}.

\begin{figure}
\centering
\includegraphics[width=0.42\textwidth,height=0.42
                 \textwidth,angle=0]{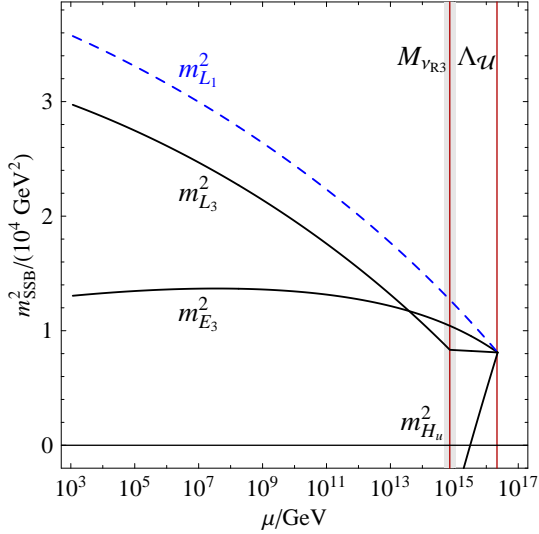}
\caption{Evolution of the first and third generation left- and right-handed
         slepton and Higgs mass parameters when
         loops involving the right-handed neutrino superfield are included;
         Ref.\,\cite{Deppisch:2007xu}.}
\label{fig15}
\end{figure}

To be specific, we focus on a simple model incorporating one-step symmetry
breaking from an SO(10) GUT model (with the Yukawa couplings of the neutrino
sector proportional to the up-type quark mass matrix) down to the SM.
Since the $\nu_R$ is unfrozen only if the RGE scale $Q$ is beyond its
mass scale $M_{\nu_R} \sim m^2_t/m_{\nu_3}\sim 7 \times 10^{14}$ GeV the impact
of the left-right extension becomes visible in the evolution of the scalar mass
parameters only at very high scales. In addition, the effect of $\nu_R$ can
be manifest only in the third generation where the Yukawa coupling is
large. In Fig.\,\ref{fig15} the evolution of the first and third
generation left- and right-handed slepton and Higgs mass parameters are displayed.
The lines include the effects of the right-handed neutrino which
induce the kink in $m^2_{L_3}$.
The kink shifts the physical masses squared of the $\tilde{\tau}_L$ and
$\tilde{\nu}_{\tau_L}$ particles of the third generation by an amount
$\Delta_{\nu_\tau}$ compared with the slepton masses of the first two
generations. The precise
measurement of $\Delta_{\nu_\tau}$ at ILC can be exploited to determine the
neutrino seesaw scale of the third generation, $M_{\nu_{R_3}}=4.7$
to $11.2\times 10^{14}$ GeV.

Before closing this section, we note that other extended scenarios with CP
violation, R-parity violation \cite{R34}, flavor violation \cite{R35},
NMSSM \cite{R36a,R36b} and/or extended gauge groups \cite{R20,R37} also are
among the paths nature may have taken. It is, therefore, strongly
recommended that the analysis conventions and methods be so general that they
can be applied to all those BSM scenarios as well.

\section{Dark matter and baryon asymmetry}
\label{sec:6}

Collider physics programs focus in connection with cosmology on two
fundamental problems \cite{Feng:2005nz,Steffen:2007sp}; the particle character of
cold dark matter (CDM), $\rho_{\rm CDM}=23\pm 4\%$, and the mechanism responsible
for the baryon asymmetry, $\rho_{\rm B}=4.0\pm 0.4\%$. These central problems cannot
be solved within the framework of the SM, but various solutions have been
worked out in the context of the BSM models.
LHC and ILC experiments are expected to play a decisive role in clarifying the nature of
CDM and in establishing the true mechanism for generating the baryon asymmetry
in the universe.

\subsection{Cold dark matter}
\label{sec:6-1}

Since there is no proper CDM particle candidate in the SM,
the presence of CDM is a clear evidence for physics beyond the SM.
In SUSY theories with R-parity the LSP is absolutely stable and represents a
good CDM candidate \cite{R8}. In particular, the lightest neutralino is
considered to be a prime candidate, but other interesting possibilities are the
the gravitino and the axino.

\begin{figure}[h]
\centering
\includegraphics[width=0.43\textwidth,height=0.43
                 \textwidth,angle=0]{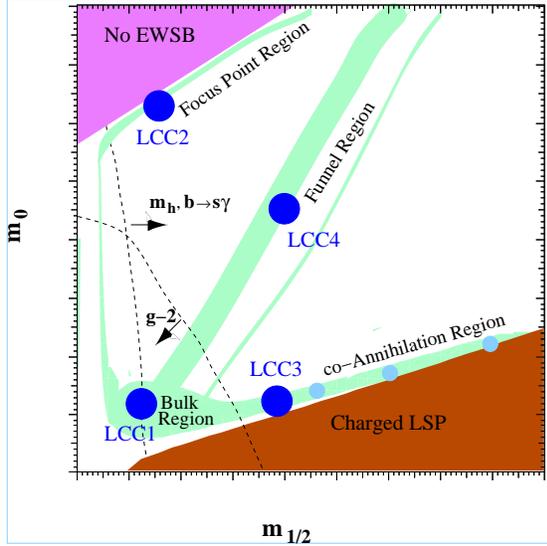}
\caption{The CDM--favored regions in the parameter space of the universal
         gaugino and sfermion mass parameters, $m_{1/2}$ and $m_0$
         with all experimental and theoretical constraints imposed;
         Ref.\,\cite{Ellis:2003cw,R38}.}
\label{fig16}
\end{figure}
\begin{figure*}[t]
\centering
\includegraphics[width=0.9\textwidth,height=0.4\textwidth,angle=0]{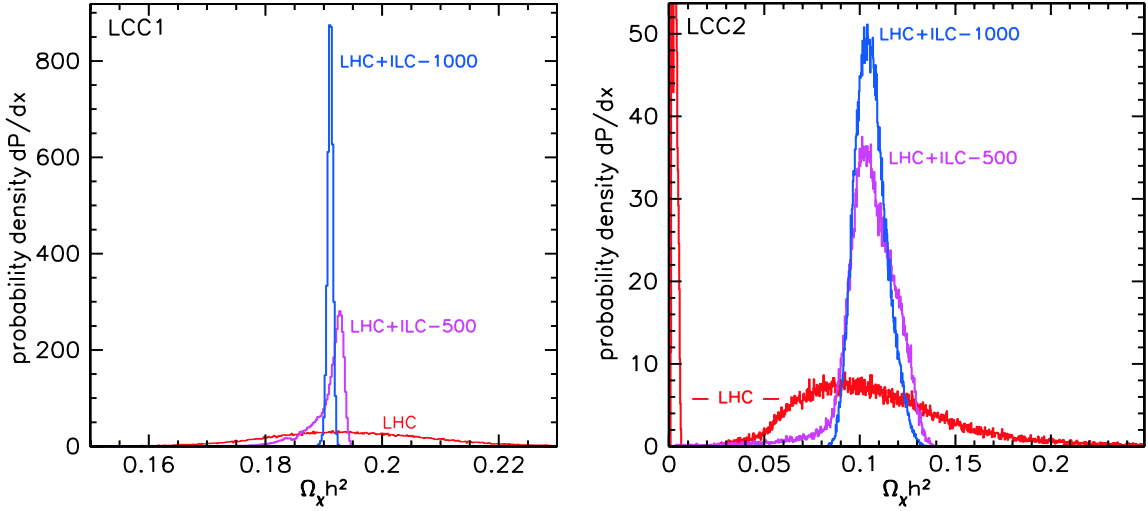}
\caption{Probability distribution of predictions for $\Omega_\chi h^2$ for
         the LCC1 [SPS1a$'$] ``bulk" point and the LCC2 ``focus-point" point from
         measurements at ILC with $\sqrt{s}=0.5$ and $1$ TeV, and LHC (after
         qualitative identification of the model); Ref.\,\cite{R38}.}
\label{fig17}
\end{figure*}

In certain areas of the SUSY parameter space with the $\tilde{\chi}^0_1$ relic
density in the range required by WMAP, SUSY particles can be produced
abundantly at LHC and ILC. However, to predict the WMAP
relic density, we must have detailed knowledge not only of the LSP
properties but also of all other particles contributing to the
LSP pair annihilation cross section. To quantify the prospects for
determining the neutralino CDM relic density at ILC as well as LHC,
four benchmark mSUGRA scenarios
called LCC points and compatible with WMAP data have been
proposed, cf. Fig.\,\ref{fig16} \cite{R38}.
The ILC measurements at $\sqrt{s}=0.5$ TeV and $1$ TeV for various
sparticle masses and mixings in the scenarios, taking into account
LHC data, are compared to those which can be obtained using LHC data
(after a qualitative identification of the model). As can be seen in
Fig.\,\ref{fig17} for two LCC points, the LCC1 ``bulk" point,
close to SPS1a$'$, and the LCC2
``focus-point" point, the gain in sensitivity by combining LHC and ILC
is spectacular.

In supergravity models the gravitino $\tilde{G}$ itself may be the LSP,
building up the dominant CDM component \cite{R39a,R39b,R39c,R39d}.
In such a scenario, with a gravitino mass in the range of 100 GeV, the lifetime
of the next-to-LSP (NLSP) can become macroscopic as the gravitino coupling is
only of gravitational strength. Special experimental efforts  are needed to catch
the long-lived $\tilde{\tau}$'s and to measure their
lifetime \cite{R40a,R40b,R40c,R40d}. Tau slepton pair production at ILC
determines the $\tilde{\tau}$ mass and the observation of the
$\tau$ energy in the $\tilde{\tau}$ decay determines the gravitino mass.
The measurement of the lifetime can subsequently be exploited to
determine the Planck scale, a unique opportunity in a laboratory
experiment.

\subsection{Baryon asymmetry}
\label{sec:6-2}

Two approaches for generating the baryon asymmetry are widely discussed in
the literature: baryogenesis mediated by leptogenesis and EW baryogenesis
based on the supersymmetric extension of the SM.

\subsubsection{Leptogenesis}

If leptogenesis is the origin of the observed baryon asymmetry, the roots
of this phenomenon are located near the Planck scale \cite{Fukugita:1986hr}.
CP-violating decays
of heavy right-handed Majorana neutrinos generate a lepton asymmetry
which is transferred to the quark sector by sphaleron processes.
Heavy neutrino mass scales as introduced in the seesaw mechanism for
generating light neutrino masses and the size of the light neutrino
masses define a self-consistent frame which
is compatible with all experimental observations \cite{Buchmuller:2004nz}.

As shown previously, in some SUSY models the size of the heavy seesaw
scales can be related to the values of the charged and neutral slepton
masses. The excellent resolution of ILC in measuring the slepton masses
can then be used to estimate the GUT-scale mass of the heaviest right-handed neutrino
within a factor of two.

\subsubsection{EW baryogenesis}

One of the conditions for generating the baryon asymmetry of the
universe requires the departure from thermal equilibrium. If triggered by
sphaleron processes at the EW phase transition, the transition must be
sufficiently strong of first order. Given the present bounds on the Higgs
mass, this cannot be realized in the SM. However, since top and stop
fields modify the Higgs potential strongly through radiative corrections,
SUSY scenarios can give rise to first-order transitions,
cf. Ref.\,\cite{Carena:1996wj}. The parameter
space of the MSSM is tightly constrained in this case: The mass of the
light Higgs boson is bounded by 120 GeV from above, and the mass of the
light stop quark is required to be smaller than the top quark mass.

This scenario suggests that the mass of the stop quark is only slightly
less than the lightest neutralino mass. The correct CDM density is
generated by stop-neutralino co-annihilation in this region of parameter
space, leading to tight constraints for the masses of the two particles.

While studies of the light stop quark are very difficult at hadron colliders
if the main decay channel is the two-body decay $\tilde{t}_1\to c
\tilde{\chi}^0_1$ with a low-energy charm jet in the final state, the clean
ILC environment allows for precision studies of the system also in
such configurations \cite{Carena:2005gc,Sopczak:2007cy}.

\section{Conclusions}
\label{sec:7}

The next generation of high energy experiments, LHC and ILC (and also CLIC later),
will usher us into the Terascale, opening a new territory which is expected
to generate a wealth of ground-breaking discoveries. The physics programme of both LHC and
ILC in exploring this microscopic world will be very rich, with unique
characteristics depending on the BSM physics scenario realized in nature.
Furthermore, as demonstrated by dedicated studies using the SUSY models, the
physics potential of LHC and ILC can significantly be extended by coherent or/and
``concurrent" running of both machines.

In summary, the LHC and ILC experiments with different advantages and
capabilities can contribute coherently and complementarily to solutions of
key questions in particle physics and cosmology. Both experiments
can eventually provide us with a comprehensive and high-resolution picture not only of
SUSY but also of any alternative scenario, serving as a telescope to
unification scenarios of matter and interactions, and connecting particle physics
and cosmology.

\subsection*{Acknowledgements}
\label{sec:ack}

Special thanks go to P.~M.~Zerwas for the critical reading of the manuscript.
This work was supported in part by the Korea Research Foundation Grant funded by
the Korean Government (MOEHRD, Basic Research Promotion Fund)
(KRF-2008-521-C00069) and in part by KOSEF through CHEP at Kyungpook National
University.

\end{document}